\documentclass[aps,pra,twocolumn,a4paper,superscriptaddress]{revtex4}

\pacs{03.67.Ac, 03.67.Lx, 42.50.Ex, 05.40.Fb}

\usepackage{amsmath}
\usepackage{amssymb}

\usepackage{color,epsfig,graphicx}

\newcommand{\ket}[1]{|#1\rangle}

\newcommand{\braket}[2] {\left< \left. #1\vphantom{#2} \right| #2 \right>} 
\newcommand{\ketbra}[2] {| #1 \rangle \! \langle #2 |}

\def\Id{\openone}

\def\tdelta[#1]{\ensuremath{\tilde{\delta}_{#1}}}

\def\xtg{\ensuremath{\vec{x}_{\mathrm{tg}}}}

\def\xm{\ensuremath{\vec{x}_{\mathrm{m}}}}
\def\tf{\ensuremath{t_{\mathrm{f}}}}
\def\Or{\mathrm{O}}

\begin{document}

\title{Optimized quantum random-walk search algorithms on the hypercube}

\author{V. Poto{\v c}ek}
\affiliation{
Department of Physics, FJFI {\v C}VUT, {B\v r}ehov\'a 7, 115
  19 Praha 1 - Star{\'e} M{\v e}sto, Czech Republic
}

\author{A. G\'abris}
\affiliation{
Department of Physics, FJFI {\v C}VUT, {B\v r}ehov\'a 7, 115
  19 Praha 1 - Star{\'e} M{\v e}sto, Czech Republic
}
\affiliation{
Research Institute for Solid State Physics and Optics, Hungarian
Academy of Sciences, H-1525 Budapest, P. O. Box 49, Hungary
}

\author{T. Kiss}
\affiliation{
Research Institute for Solid State Physics and Optics, Hungarian
Academy of Sciences, H-1525 Budapest, P. O. Box 49, Hungary
}

\author{I. Jex}
\affiliation{
Department of Physics, FJFI {\v C}VUT, {B\v r}ehov\'a 7, 115
  19 Praha 1 - Star{\'e} M{\v e}sto, Czech Republic
}

\date{\today}
\begin{abstract}
  Shenvi, Kempe and Whaley's quantum random-walk search (SKW)
  algorithm [Phys.\ Rev.\ A \textbf{67}, 052307 (2003)] is known to
  require $O(\sqrt N)$ number of oracle queries to find the marked
  element, where $N$ is the size of the search space. The overall time
  complexity of the SKW algorithm differs from the best achievable on
  a quantum computer only by a constant factor. We present
  improvements to the SKW algorithm which yield significant increase
  in success probability, and an improvement on query complexity such
  that the theoretical limit of a search algorithm succeeding with
  probability close to one is reached. We point out which improvement
  can be applied if there is more than one marked element to find.
\end{abstract}

\maketitle

\section{Introduction}

In the pioneering paper \cite{shenvi:052307} Shenvi, Kempe and Whaley
demonstrated that a useful quantum algorithm can be designed based on
quantum random-walks. This quantum random-walk search algorithm (the
SKW algorithm) can be used to find a vertex of a hypercube that is
marked by an oracle.  Although the number of oracle calls needed by
the SKW algorithm scales with the size of the search space similarly
to the Grover search \cite{Grover}, its principle of operation is
significantly different.  Since the pioneering work a variety of
quantum algorithms have been proposed utilizing quantum random-walks,
see for example \cite{AmbainisQuantum-search-,Santha_TAMC08}.  The SKW
algorithm may be divided into a quantum part, and a simple classical
protocol in which the former is embedded. The quantum part is a
perturbed Grover walk on a hypercube started from an equally weighted
superposition of initial states and iterated for a given number of
steps, to be followed by a measurement on the output state to find the
marked vertex. The perturbation of the Grover coin is derived from the
oracle, which is used to introduce position dependence into the coin
operator. In this paper, we shall use the term SKW quantum random-walk
to refer to this special quantum random-walk. As it has been shown in
\cite{shenvi:052307} the SKW quantum random-walk yields the marked
vertex with probability strictly less than $1/2$, therefore it is
necessary to embed it into a classical protocol to find the marked
vertex with certainty, or use an amplitude amplification scheme
\cite{Brassard:qph.970402, grover:prl.80.4329}. The classical protocol
of the SKW algorithm is relatively simple: a measurement is made on
the final state of the SKW quantum random-walk, then its result is
verified by querying the oracle directly. By repeating the algorithm
and these two steps a sufficient number of times, we can make sure
that the marked element is found with an arbitrary small failure
probability. Applying an amplitude amplification scheme would provide
a more efficient way for increasing success probability, however, its
use would mean departure from the quantum random walk paradigm.

The overhead caused by repeating the quantum random-walk several
times, although contributing only a constant factor to the time
complexity, can be a considerable source of difficulties in certain
experimental scenarios. In the present paper we present modifications
to the SKW algorithm which allow significant reduction of the number
of necessary repetitions. We note that in 2 dimensions the spatial
search algorithm by Ambainis, Kempe and Rivosh
\cite{Ambainis:0402107} also yields the target vertex after one run
with probability less than one, i.e.\ with only $\Theta(1/\sqrt{\ln
  N})$. Recently, Tulsi \cite{Tulsi:pra78.012310} has proposed
improvements to this algorithm, which allow the finding of the target
vertex with probability $1$ after one run. The speedup in
\cite{Tulsi:pra78.012310} has been achieved by introducing an ancilla
qubit into the computational space, which is similar in spirit to our
improvement modification described in Sec.~\ref{sec:skw-optimal} which
uses an additional coin dimension. Improvements of quantum walk-based
searches have been studied also by other authors. In
\cite{Reitzner:qph0805.1237} an optimization dedicated to the
scattering random-walk implementation \cite{hillery:032314,
  Kosik2005Scattering-mode, gabris:pra_76_062315} has been proposed,
related to the findings we describe in
section~\ref{sec:prob-neighbour}. In
\cite{CMChandrashekar:pra77_032326} the authors discussed the
optimization of the quantum walk on a line by varying the coin
operator parameters.

In Section~\ref{sec:prob-neighbour} we prove that the final state of
the SKW quantum walk consists mainly of the target vertex and its next
neighbours, and present modifications to the algorithm which exploit
this property. These modifications can be used to reduce the number of
repetitions of the SKW quantum walk, and to reduce the number of
independent verification queries to the oracle. We note that the task
of verification may be problematic for certain implementations, e.g.\
in a spatial search implementation where a vertex being marked is a
local property and not a property given by an oracle. Such additional
costs have been considered in Ref.~\cite{Magniez-acm39} in connection
with quantum walks.

Based on the SKW algorithm we develop an algorithm in
Section~\ref{sec:skw-optimal} that displays query complexity $1/\sqrt
2$ of the original, thus the theoretically lowest for a search
algorithm with a success probability close to 1
\cite{Zalka-pra:60.2746}. Our improvement is founded on the bipartite
nature of the SKW quantum random-walk, and we arrive at its final form
after several steps. We note that some of these intermediate steps may
be useful improvements on their own right, depending on the actual
physical implementation.

In Section~\ref{sec:multiple-marked} we outline the conditions under
which the optimizations introduced in Section~\ref{sec:skw-optimal}
can be used to find multiple marked vertices. Finally, in
section~\ref{sec:conclusions} we conclude our results.

\section{Improving success probability by considering next neighbours}
\label{sec:prob-neighbour}

In this section we describe a property of the SKW quantum walk that
can be used to boost the probability of finding the marked vertex by
doing a proper measurement on its final state. Let
$\mathcal{C}_n=(V_n, E_n)$ denote the graph of the $n$ dimensional
hypercube. The argumentation of the present paper relies heavily on
the concept of the Hamming weight and the parity of an integer, which
can be easily related to each other. The Hamming weight of an integer
is the number of 1s in its binary string representation $\vec{x}$, and
shall be denoted by $|\vec{x}|$ in this paper. The parity of $\vec{x}$
is then simply $|\vec{x}| \mathop{\mathrm{mod}} 2$. A related concept
is the Hamming distance of two integers, say $\vec{x}$ and $\vec{y}$,
that is defined as $|\vec{x} \oplus \vec{y}|$, where $\oplus$ denotes
the bitwise addition modulo $2$ operator. Following the notation of
earlier work \cite{shenvi:052307, moore02quantum}, the vertices $V_n$
of the hypercube are labelled by integers $\vec{x} = 0, \ldots, 2^n-1$
in such a way that the Hamming distance between any two vertices
connected by an edge is exactly 1. The SKW quantum walk takes place on
the product Hilbert space $\mathcal{H}^{C_n} \otimes
\mathcal{H}^{V_n}$ where $\mathcal{H}^{V_n}$ is the $N=2^n$
dimensional Hilbert space representing the vertices, and
$\mathcal{H}^{C_n}$ is the $n$ dimensional space associated with the
quantum coin. The propagator of the SKW quantum walk can therefore be
written as
\begin{equation}
S = \sum_{d,\vec{x}} \ketbra{d,\vec{x} \oplus \vec{e}_d}{d,\vec{x}}
\label{eq:full-S},
\end{equation}
where $\vec{e}_d=2^d$ correspond to the edges originating from the
given vertex. If the target vertex marked by the oracle
$\mathcal{O}$ is denoted by $\xtg$, the perturbed coin operator can
be written as
\begin{equation}
C' = C_0 \otimes \Id + (C_1 - C_0) \otimes \ketbra{\xtg}{
\xtg}.
\label{eq:full-C}
\end{equation}
For the SKW quantum walk, $C_0$ is usually chosen to be the $n$
dimensional Grover operator (also known as the Grover diffusion
operator) and $C_1$ is chosen to be $-\Id$. The results in this
section, however, hold for any pair of inequivalent permutation
invariant unitary coins. As it has been argued in
\cite{shenvi:052307}, due to the symmetry of the hypercube graph the
vertices can always be re-labelled in such a way that the marked
vertex becomes $\xtg=0$. Since with this choice the permutation
invariance of the Grover walk on the hypercube is conserved, the
initial state
\begin{equation}
  \ket{\psi_0} = \frac1{\sqrt{n 2^n}} \sum_{d=1}^n \sum_{\vec{x}}
  \ket{d,x}
  \label{eq:psi0}
\end{equation}
allows the reduction to a walk on a line. The basis states for this
collapsed quantum walk are defined as
\begin{eqnarray}
\ket{R,x} &=& \sqrt{\frac{1}{(n-x) {n \choose x} }} \sum_{|\vec{x}|=x}
\sum_{x_d=0} \ket{d,\vec{x}}, \label{eq:R-basis} \\
\ket{L,x} &=& \sqrt{\frac{1}{x {n \choose x} }} \sum_{|\vec{x}|=x}
\sum_{x_d=1} \ket{d,\vec{x}},
\label{eq:L-basis}
\end{eqnarray}
and the propagator becomes
\begin{equation}
  \label{eq:S}
  S= \sum_{x=0}^{n-1} \ketbra{R,x}{L,x+1} + \ketbra{L, x+1}{R,x}.
\end{equation}
The coin operator of the walk on the line acquires a strong position
dependence. For example when $C_0$ is the Grover coin, in the
collapsed basis it becomes
\begin{equation}
C_0 = \sum_{x=0}^n \left( \begin{array}{rr} \cos \omega_x & \sin
    \omega_x \\ \sin \omega_x & - \cos \omega_x \end{array} \right)
\otimes \ketbra{x}{x},
\end{equation}
where $\cos \omega_x = 1 - 2x/n$ and $\sin \omega_x = (2/n) \sqrt{x
  (n-x)}$, and the matrix is understood in the $\{ \ket{R}, \ket{L}
\}$ basis. The perturbed coin with $C_1=-\Id$ can be written as
\begin{equation}
  C' = C_0 -  2 \ketbra{R,0}{R,0}.
\end{equation}

\begin{figure}
\begin{center}
\includegraphics[width=.45\textwidth]{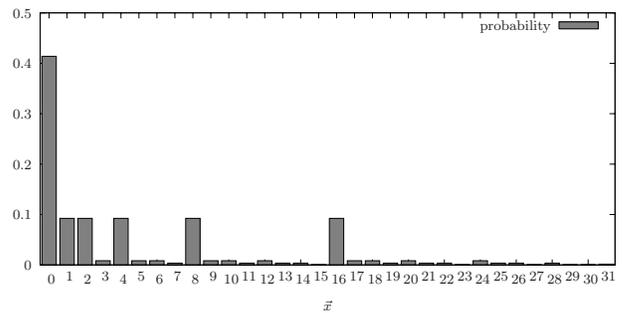}
\end{center}
\caption{The plot shows the numerically calculated probability
  distribution for the position of the walker after the optimal number
  of iterations of the SKW quantum walk in $n=5$ dimensions. In
  accordance with the analytic results, the probability distribution
  has its maximum for the marked vertex $\xtg=0$ reaching a value
  close to 1/2.  Moreover, we observe that the nearest neighbours are
  also presented with high probability, and the sum of these
  probabilities is comparable to that of the marked vertex. }
\label{fig:prob-condense}
\end{figure}

It has been shown in \cite{shenvi:052307} that after an optimal number
of iterations the probability $p_0$ of obtaining the target state
$\ket{0}$ in a measurement is close to $1/2$, and that the optimal
number of iterations is well estimated by the nearest integer to
\begin{equation}
  \tf = (\pi/2) \sqrt{2^{n-1}}.
  \label{eq:tf}
\end{equation}
This means that the final state is composed mainly of the target
state, and contains smaller contributions from its next and more
distant neighbours \cite{shenvi:052307}. However, this statement can
be refined by partitioning the SKW quantum walk into two independent
quantum walks. Let $\mathcal{H}_e$ denote the subspace spanned by
states $\ket{d,\vec{x}}$ such that $|\vec{x}|$ is even, and
$\mathcal{H}_o$ denote the subspace spanned by the states with
$|\vec{x}|$ being odd. The terms even and odd refer to a labelling
where the target vertex is denoted by $\xtg=0$, therefore, in general,
these subspaces must be defined according to the parity of
$\vec{x}\oplus \xtg$.  The two quantum walks are started in the
Hilbert spaces $\mathcal{H}_e$ and $\mathcal{H}_o$, and evolve
independently. In the following we shall term $\mathcal{H}_e$ the
\textit{even} subspace, and $\mathcal{H}_o$ the \textit{odd} subspace
of $\mathcal{H}$. It follows from the property of the parity function
that this partitioning of $\mathcal{H}$ is the same for all values of
$\xtg$, however, the role of the two subspaces depends on the parity
of $\xtg$. We can define the orthogonal projectors $P_e$ and $P_o$
that project to $\mathcal{H}_e$ and $\mathcal{H}_o$, respectively.
Clearly, in the collapsed basis, the \textit{even} subspace is spanned
by the states (\ref{eq:R-basis}) and (\ref{eq:L-basis}) with $x$ being
even, and the \textit{odd} subspace is spanned by those with $x$ being
odd. Since $ [P_{e/o}, C_0] =0$ and $[P_{e/o}, C'] = 0$ it follows
from the definition of $S$ that
\begin{subequations}
\label{eq:U-ss-switch}
\begin{eqnarray}
  P_o U' &=& U' P_e, \\
  P_e U' &=& U' P_o. 
\end{eqnarray}
\end{subequations}
Let us introduce the (normalized) states
\begin{eqnarray}
  \ket{\psi^{(e)}_0} &=& \sqrt2 P_e \ket{\psi_0}, \label{eq:psi-e0}\\
  \ket{\psi^{(o)}_0} &=& \sqrt2 P_o \ket{\psi_0}, \label{eq:psi-o0}
\end{eqnarray}
and express the initial state as $\ket{\psi_0} = \frac1{\sqrt2} (
\ket{\psi^{(e)}_0} + \ket{\psi^{(o)}_0})$. It can easily be seen that
the action of $U'$ on $\ket{\psi^{(o)}_0}$ simplifies to
\begin{equation}
  \label{eq:U-oddpsi}
  U' \ket{\psi^{(o)}_0} = U \ket{\psi^{(o)}_0} = \ket{\psi^{(e)}_0}.
\end{equation}
By successive applications of Eqs.~(\ref{eq:U-ss-switch}) and
(\ref{eq:U-oddpsi}) it can be shown that $U'$ has the property
\begin{subequations}
  \label{eq:even-odd-steps}
\begin{eqnarray}
  P_e (U')^{2r} \ket{\psi_0} &=& P_e (U')^{2r+1} \ket{\psi_0}, \quad
  (r=0, 1, 2, \ldots), \label{eq:even-steps} \\
  P_o (U')^{2r} \ket{\psi_0} &=& P_o (U')^{2r-1} \ket{\psi_0}, \quad
  (r=1, 2, 3, \ldots), \label{eq:odd-steps} 
\end{eqnarray}
\end{subequations}
Let us express the state of the walker after $t$ steps as
\begin{equation}
(U')^t \ket{\psi_0} = \sum_{x=0}^{n-1} \alpha_{R,x}^t \ket{R,x} +
\sum_{x=1}^{n} \alpha_{L,x}^t \ket{L, x}  ,
\label{eq:alpha-notation}
\end{equation}
and define $P_x^t = |\alpha_{L,x}^t|^2 + |\alpha_{R,x}^t|^2$, with
setting $\alpha_{R,n}^t=\alpha_{L,0}^t=0$ for convenience. The
interpretation of $P_x^t$ is clear from the definitions: $P_0^t$ is
the probability of having the walker at the target vertex after $t$
iterations, and $P_1^t$ is the total probability of finding the walker
at any of the nearest neighbours of the target node. Using the above
bipartition of the quantum walk it can be shown that the inequalities
\begin{subequations}
  \label{eq:rel-P0-P1}
  \begin{eqnarray}
    P^{t+1}_0 &\le& P^t_1 \label{eq:rel-P0-P1-a},\\
    P^{t-1}_0 &\le &P^t_1 \label{eq:rel-P0-P1-b}
  \end{eqnarray}
\end{subequations}
hold for all $t>0$. From which it follows that the probabilities
\begin{eqnarray}
p_0 &=& \sum_{d=0}^n |\braket{d,0}{\psi_{\mathrm{f}}}|^2, \\
p_1 &=& \sum_{d,|\vec{x}|=1}^n |\braket{d,\vec{x}}{\psi_{\mathrm{f}}}|^2,
\label{eq:prob_0_1}
\end{eqnarray}
satisfy the inequality
\begin{equation}
p_1 \ge p_0.
\label{eq:p1-ge-p0}
\end{equation}
For the details of the calculations see Appendix~\ref{app:a}. This
property is illustrated in Fig.~\ref{fig:prob-condense} using a
numerical simulation.

Therefore, since we have $p_0=1/2 - \Or(1/n)$, the total probability
of measuring the target node or any of its direct neighbours is
\begin{equation}
\label{eq:condensed-prob}
p_c = p_0 + p_1 \ge 1 - \Or(1/n).
\end{equation}
Since $p_c$ is upper bounded by $1$, for large $n$ the total
probability must be approaching $p_c = 1$. Naturally, the question
arises: can Eq.~(\ref{eq:condensed-prob}) be turned to our advantage?
In the following we shall address this question, and answer
positively.

First, let us analyze the most straight-forward way of taking
advantage of Eq.~(\ref{eq:condensed-prob}). According to the SKW
protocol, the validity of the measurement outcome $\xm$ after $\tf$
iterations is verified using the oracle. If the verification is
positive the target node is found, otherwise the result is discarded
and the SKW quantum walk is repeated. However, this is unnecessary
since from Eq.~(\ref{eq:condensed-prob}) we know that in case of a
negative answer from the oracle, the probability that $\xm$ is a
direct neighbour of $\xtg$ is greater than $1 - \Or(1/n)$. Therefore,
it is sufficient to query the oracle with values from the set $\{ \xm
\oplus \vec{e}_d \mathop{|} d=0,\ldots,n-1 \}$, from which the marked
element can be extracted using the simplest classical protocol by an
average of $(\log_2 N)/2$ additional oracle queries. 

In a scenario where the verification costs are dominating over all
other cost it is crucial to perform the minimum number of necessary
verification queries. One possibility could be to use amplitude
amplification or another quantum based search, however, both these
approaches mean a departure from the original hypercube quantum
random-walk.

In the following, we propose an alternative approach to reduce the
number of verification queries if the coin states can also be
determined. Let us set $t_{f,o} = 2 \lfloor \tf/2 \rfloor + 1$, and
denote the outcome of the measurement on the coin state by $d_{\mathrm
  m}$.  Using the notations of (\ref{eq:alpha-notation}), we can
re-write Eq.~(\ref{eq:even-steps}) for the case $j=0$, and obtain
$|\alpha_{R,0}^{t_o-1}|^2 = |\alpha_{R,0}^{t_o}|^2 = 1/2 - \Or(1/n)$.
From the unitarity of the coins, and the definition (\ref{eq:S}) of
$S$ it follows that we also have
\begin{equation}
  |\alpha_{L,1}^{t_o}|^2 = |\alpha_{R,0}^{t_o}|^2 = \frac12 - \Or(1/n).
\end{equation}
Note, that this ensures also that he have $\alpha^{t_o}_{R,1} =
\Or(1/n)$, which is negligible for large $n$. Therefore, we conclude
that the final state is composed mainly of the states $\ket{R,0}$ and
$\ket{L,1} = 1/\sqrt{n} \sum_{d=0}^{n-1} \ket{d, \vec{e}_d}$. Thus, if
the measurement gives $\xm \neq \xtg$ then the target vertex can be
found with $1-\Or(1/n)$ probability by taking $\xtg = \xm \oplus
\vec{e}_{d_{\mathrm m}}$.

In other words, if a complete measurement can be made on the coin
state, the marked element can be determined with $1-\Or(1/n)$
probability after a single execution of the SKW algorithm and one
verification query to the oracle.

\section{Modification to attain optimal query complexity}
\label{sec:skw-optimal}

In the present section, based on the SKW algorithm we develop a search
algorithm which finds the marked vertex of a hypercube using the
optimal number of oracle queries. In contrast to the modifications of
\ref{sec:prob-neighbour} which essentially affect only the classical
processing part, the improvement proposed in the present section
requires a modification of the quantum walk itself.

The improvement is based on the bipartite nature of the SKW quantum
walk, which implies the invariance of the \textit{even} and
\textit{odd} subspaces under two iterations of $U'$,
\begin{equation}
  [P_o, U^{\prime 2r}] = [P_e, U^{\prime 2r}] = 0, \quad (r = 0, 1,
  \ldots),
\label{eq:e-o-invariance}
\end{equation}
which follows from Eqs.~(\ref{eq:U-ss-switch}). First, consider the
projection of the state of the walker after $2r$ iterations onto the
\textit{even} subspace. In the spirit of Eq.~(\ref{eq:e-o-invariance})
we can see that the projection of the final state corresponds to a similar
projection of the initial state, which we can write as
\begin{equation}
  U^{\prime 2r} \ket{\psi_0^{(e)}} = \sqrt2 P_e U^{\prime 2r}
  \ket{\psi_0}.
\end{equation}
Introducing $t_{f,e} = 2 \lfloor \tf/2 \rfloor$ we conclude that
for the probability $P^{(e) t_{f,e}}_0$ to find the marked node after
$t_{f,e}$ iterations starting from the \textit{even} initial state
$\ket{\psi^{(e)}_0}$ the relation
\begin{equation}
P^{(e) t_{f,e}}_0 = 2 P^{t_{f,e}}_0 = 1 - \Or(1/n)
\label{eq:psie0-double}
\end{equation}
holds. This is an encouraging result, since it suggests that the
marked element can be directly found with high probability after a
single execution of the SKW algorithm without any verification
queries.  However, the choice $\xtg=0$ is actually the result of the
mapping $\vec{x} \to \vec{x} \oplus \xtg$, thus we do not know in
general which is the \textit{even} subspace and which is the
\textit{odd} subspace.

The information about the parity of the marked vertex is clearly
contained in the oracle. An efficient way of extracting this
information is to repeat the quantum walk twice, once starting from
the initial state $\ket{\psi^{(e)}_0}$ and once starting from
$\ket{\psi^{(o)}_0}$. Note, that it is not necessary to know which
one is which, since $\ket{\psi^{(e)}_0}$ will yield $\ket{\xtg}$ with
nearly unit probability. Therefore, the target vertex can be
identified by testing the two measurement outcomes
$x^{(e)}_{\mathrm{m}}$ and $x^{(o)}_{\mathrm{m}}$ on the oracle.

Instead of repeating the algorithm twice, it is possible to construct
another SKW quantum walk in which it is guaranteed that the marked
element corresponds to a vertex with even parity. The principle of
this modification is the mapping of all the vertices of the $n$
dimensional hypercube to the even parity vertices of an $n'=n+1$
dimensional hypercube. Since the number of even and odd vertices is
equal for a hypercube in every dimension, the mapping between the
original vertices and the even parity vertices of the larger hypercube
can be made one-to-one.

In the following, we assume that the oracle is given as an operator
acting on the Hilbert space $\mathcal{H}^{V_n}$ associated to the $n$
dimensional hypercube, and we shall construct an SKW quantum walk in
$n'=n+1$ dimensions using the extended oracle acting on the Hilbert
space $\mathcal{H}^{V_{n'}}$. The vertices $\vec{x}$ of the original
hypercube are mapped to the even parity sites of the extended
hypercube by the map
\begin{equation}
  \vec{x}' = m(\vec{x}) = 2\vec{x} + p(\vec{x}), \label{eq:map}
\end{equation}
where $p(\vec{x})$ denotes the parity of $\vec{x}$. This mapping can
be viewed as appending one bit to the bit string representation of the
original vertex, the value of the bit being $1$ for odd parity
vertices, and $0$ for even parity vertices. The reverse mapping simply
drops the appended bit for even parity input, while the odd parity
vertices of the extended hypercube do not correspond to any vertices
of the original graph. 

\begin{figure}
\begin{center}
\includegraphics{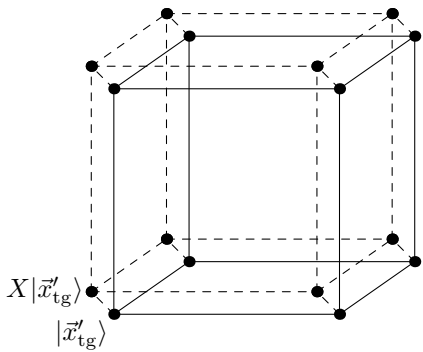}
\end{center}
\caption{Extension of the search in $n=3$ dimensions to a hypercube in
  $n'=4$ dimensions without distinguishing even and odd parity
  vertices. The operator $X$ denotes the application of the Pauli
  $\sigma_X$ operation to the last qubit. The effect of operator $X$
  is switching between the image and the anti-image of a vertex.}
\label{fig:double-hypercube}
\end{figure}

In this way the marked vertex is known to be mapped to an even parity
vertex on the extended hypercube. The modification of the oracle
$\mathcal{O}$ to return positive result only for the new marked vertex
is straight forward.  Let the operators of the $n'$ dimensional
extended SKW quantum walk be distinguished from the original $n$
dimensional one by adding a $(+)$ superscript. Therefore, the coin
operators acting on $\mathcal{H}^{C_{n'}}$ are denoted by $C_0^{(+)}$
and $C_1^{(+)}$, and the propagator operator on $\mathcal{H}^{C_{n'}}
\otimes \mathcal{H}^{V_{n'}}$ by $S^{(+)}$. Similarly, the perturbed
coin operator is denoted by $C'^{(+)}$. With this mapping, the
procedure described above can be applied very efficiently since the
``good'' initial state $\ket{\psi^{(e)}_0}$ is prescribed by the
construction. Consequently, a single execution of the $n'$ dimensional
SKW quantum walk is sufficient to find the marked vertex with a
probability close to unity. Note that the extension to $n'=n+1$
dimensions changes the optimal number of iterations, which amounts to
an increase of the query complexity by a factor of $\sqrt{2}$.

The query complexity can be reduced by noting that at every second
iteration, the coin operator $C^{(+)}$ could effectively be replaced
by the unperturbed coin operator $C^{(+)}_0$, thus the number of
oracle queries can be reduced by $1/2$. Moreover, as shown in
Appendix~\ref{app:b}, by forcing the coin operator to be $C^{(+)}_0$
for every second iteration, the equality
\begin{equation}
  (U^{(+)}U''^{(+)})^r \ket{\psi_0} = \frac1{\sqrt2} (X+\Id)
  (U^{(+)}U''^{(+)})^r \ket{\psi_0^{(e)}}
\end{equation}
holds, where $X$ denotes the quantum \textsc{not} gate, $\sigma_X$,
acting on the last qubit.  Thus, an initial state of uniform
superposition (\ref{eq:psi0}) can be used, yielding the image
($\xtg'$) and the anti-image ($\xtg' \oplus 1$) of the target
vertex with a total probability close to one. Therefore, by performing
a measurement that ignores the last qubit we obtain the marked vertex
$\xtg$ with probability $1 - \Or(1/n')$.

Using the formula (\ref{eq:tf}) to calculate the query complexity, we
find that the modified algorithm completes using $\tf' = (\pi/4)
\sqrt{N}$ oracle queries which is identical to what is needed by the
Grover search algorithm, and known to be the best achievable on a
quantum computer for a success probability of one
\cite{Zalka-pra:60.2746}.

The storage complexity of the improved algorithm can be reduced by
noting that the auxiliary qubit can be eliminated using the identities
\begin{eqnarray}
  [X, U^\mathrm{(+)}] &=& [X, U''{}^\mathrm{(+)}] = 0,\\
  X \ket{\psi_0} &=& \ket{\psi_0}.
\end{eqnarray}
Clearly, the reduction affects only the dimensionality of the position
space, and leaves the coin space $n'=n+1$ dimensional. With some
algebra, we obtain the reduced propagator from $S^\mathrm{(+)}$ as
\begin{equation}
  \tilde{S} = \sum_{\vec x} \left( \sum_{d=0}^{n-1} \ketbra{d,\vec x
      \oplus \vec e_d}{d,\vec x} + \ketbra{n,\vec x}{n,\vec x} \right).
\end{equation}
Thus, the coin states $\ket{d}$ with $d<n$ become the coin states of a
quantum random-walk on the original $n$ dimensional hypercube, while
the state $\ket{n}$ corresponds to a coin state instructing the walker
to remain at the same vertex at the next iteration.

The propagator can equivalently be understood as describing a quantum
random-walk on a regular graph consisting of an $n$ dimensional
hypercube having a self loop edge attached to each of its vertices.
The final version of the quantum walk for optimal search can therefore
be expressed by the alternating sequence of the unitary operators
\begin{eqnarray}
  \tilde{U}'' &=& \tilde{S} C^{\prime\prime (+)}, \\
  \tilde{U} &=& \tilde{S} C^{(+)}_0,
\end{eqnarray}
acting on an $N=2^n$ dimensional vertex space, and an $n+1$
dimensional coin space.

\section{Applications to finding multiple marked vertices}
\label{sec:multiple-marked}

In the present section we consider the optimization problem when the
number of marked vertices is more than one. Although the SKW algorithm
is guaranteed to work only when the oracle marks a single vertex,
numerical calculations suggest that it can also be used to find
multiple marked vertices as long as the number of marked vertices is
small compared to the size of the search space. 

To answer the question whether the SKW algorithm can be used to find
multiple marked vertices is beyond the scope of the present paper.
Instead, here we focus on the question of applicability of the
improvement described in Sec.~\ref{sec:skw-optimal}. In the following,
we shall show that the modified algorithm can be applied directly to
the search for multiple marked vertices when the SKW algorithm on the
extended hypercube yields sufficient results. To formalize the task of
finding multiple marked vertices, let us denote the number of elements
marked by the oracle by $m$, and their labels by $\xtg^{(j)}$, such
that $j=1, \ldots, m$. The coin operator of the SKW quantum walk can
therefore be written as
\begin{equation}
 C'_m = C_0 \otimes \Id + (C_1-C_0) \otimes
 \sum_{j=1}^m \ketbra{\xtg^{(j)}}{\xtg^{(j)}},
\label{eq:mult-coin}
\end{equation}
and the unitary evolution operator as $U'_m = S C'_m$. This unitary
operator is then iterated a given number of times to obtain a final
state that is composed mainly of the states corresponding to the
marked vertices.

For simplicity, here we consider the variant of the improvement using
the walk on the extended $n'=n+1$ dimensional hypercube using the even
parity initial state. This is sufficient, since it is equivalent to
those what we obtain by using the quantum walk with two coins, and
after the reduction back to the $n$ dimensional hypercube.  Clearly,
by defining the $n'$ dimensional extension $C'^{(+)}_m$ of $C'_m$ we
arrive at the unitary evolution operator $U'^{(+)}_m$ which also obeys
\begin{equation}
\label{eq:IV-uupp}
\ket{\psi^{(e)}(2r)} = U'^{(+)2r}_m \ket{\psi_0^{(e)}} 
= (U^{(+)}_mU''^{(+)}_m)^r \ket{\psi_0^{(e)}},
\end{equation}
since all the marked vertices are mapped to the \textit{even}
subspace. For the same reason, we have for every $d$, the relation
\begin{equation}
|\langle d, \xtg^{(j)\prime} | U'^{(+)2r}_m  |\psi_0^{(e)}\rangle|^2 
= 2 |\langle d, \xtg^{(j)\prime} | U'^{(+)2r}_m  |\psi_0\rangle|^2,
\end{equation}
according to the definition (\ref{eq:psi-e0}). Therefore, if the total
probability of finding any of the marked vertices in the final state
of the extended SKW algorithm is close to $1/2$, the modified
algorithm yields them with probability close to unity.

\section{Conclusions}
\label{sec:conclusions}

We have proposed two alternative approaches for improving the SKW
quantum random-walk search algorithm. Both improvements are centered
around increasing the success probability after one run. In the first
part of the paper we shown that the next neighbours of the target can
be obtained with high probability, and that this can be exploited to
reduce the number of repetitions or independent oracle queries to one
or two. We note, that for certain implementations, a lower repetition
count may have a serious impact on efficiency. In the second part of
the paper we have developed a two-coin quantum random-walk search
algorithm on a hypercube with self-loop edges. We have pointed out
that the speedup over the original SKW algorithm in terms of oracle
queries is $1/\sqrt2$. This makes the algorithm equivalent to the
Grover search in terms of query complexity, therefore, present an
optimal solution to the search problem if the success probability of 1
is required \cite{Zalka-pra:60.2746}.

We have also considered the optimization problem of finding multiple
marked vertices. We have shown that if the SKW quantum walk mapped to
to an $n+1$ dimensional hypercube yields the marked vertices with
probability close to 1/2, the algorithm in \ref{sec:skw-optimal} can
be applied unmodified, resulting in the same improvement as for the
case of a single marked vertex.

\acknowledgments{

  This work was supported by the Czech and Hungarian Ministries of
  Education (CZ-10/2007), by MSMT LC 06002 and MSM 6840770039 and by
  the Hungarian Scientific Research Fund (T049234 and T068736). The
  authors would like to thank Dr. E. Andersson for valuable comments
  and hospitality at Herriot-Watt University. The financial support by
  Royal Society under 2006/R2 IJP is gratefully acknowledged.

}

\appendix

\section{}
\label{app:a}

Using the notation of Eq.~(\ref{eq:alpha-notation}), let us consider
an arbitrary $\alpha^{t-1}_{R,0}$ ($\alpha^{t-1}_{L,0}$ is set to 0 by
definition).  In one iteration, $\alpha^{t-1}_{R,0}$ is first
transformed to some $\beta^{t-1}_{R,0}$ and $\beta^{t-1}_{L,0}$ by the
coin operator $C'$.  Upon inspecting the definition of $\ket{R,0}$ we
find that due to the unitarity of the coin $C'$ we have
$|\beta^{t-1}_{R,0}|^2 = |\alpha^{t-1}_{R,0}|^2 = P^{t-1}_0$ and
$\beta^{t-1}_{L,0}=0$.  Considering the action of $S$ we obtain
$\alpha^{t}_{L,1} = \beta^{t-1}_{R,0}$. Therefore, we can write $P^t_1
= |\alpha^{t}_{R,1}|^2 + |\alpha^{t}_{L,1}|^2 \ge |\alpha^{t}_{L,1}|^2
= |\alpha^{t-1}_{R,0}|^2 = P^{t-1}_0$, which proves
Eq.~(\ref{eq:rel-P0-P1-a}). The second inequality can be proven along
similar lines. Due to the unitarity of the coins we always have
$|\beta^{t}_{R,1}|^2 + |\beta^{t}_{L,1}|^2 = |\alpha^{t}_{R,1}|^2 +
|\alpha^{t}_{L,1}|^2$, and according to the definition of $S$,
$\alpha^{t+1}_{R,0} = \beta^t_{L,1}$ also holds. Therefore, we can now
write $P^{t+1}_0 = |\alpha^{t}_{R,0}|^2 = |\beta^{t}_{L,1}|^2 \le
|\beta^{t}_{R,1}|^2 + |\beta^{t}_{L,1}|^2 = P^t_1$, which provides
Eq.~(\ref{eq:rel-P0-P1-b}).

From Eqs.~(\ref{eq:even-odd-steps}) follows that $P^{2r}_x =
P^{2r+1}_x$ if $x$ is even, and that $P^{2r}_x = P^{2r-1}_x$ if $x$ is
odd. Combining these equalities with Eqs.~(\ref{eq:rel-P0-P1}) we
obtain
\begin{equation}
\label{eq:P1-ge-P0}
P^t_1 \ge P^t_0,
\end{equation}
for every positive integer $t$. Eq.~(\ref{eq:p1-ge-p0}) is a special
case of Eq.~(\ref{eq:P1-ge-P0}).

\section{}
\label{app:b}

First note that $C'^{\mathrm{(+)}} P_o = (C_0^{(+)} \otimes \Id) P_o$
holds, therefore we have
\begin{equation}
\label{eq:III-UpUUp}
U'^{\mathrm{(+)}2r} \ket{\psi_0^{(e)}} = (U^{(+)}U'^{(+)})^r
\ket{\psi_0^{(e)}}, 
\end{equation}
since Eqs.~(\ref{eq:U-ss-switch}) hold for hypercubes in all
dimensions.  Moreover, we can write
\begin{equation}
\label{eq:III-CpPe}
C'^{\mathrm{(+)}} P_e = C''^{\mathrm{(+)}} P_e,
\end{equation}
by introducing
\begin{equation}
\label{eq:III-cpp}
C''^{\mathrm{(+)}} = \left[ C_0^\mathrm{(+)} \otimes \Id + (C_1^{(+)}
  - C_0^{(+)}) \otimes \ketbra{\xtg}{\xtg} \right] \otimes \Id_2,
\end{equation}
where $\Id_2$ is the identity acting on the qubit added by the
extension. We can use the coin (\ref{eq:III-cpp}) to define the
unitary evolution operator $U''^{\mathrm{(+)}} = S^\mathrm{(+)}
C''^{\mathrm{(+)}}$. By considering the expression that gives the
final state of the walker after $2r$ steps we find that it can be
simplified to
\begin{equation}
\label{eq:III-uupp}
U'^{(+)2r} \ket{\psi_0^{(e)}} = (U^{(+)}U''^{(+)})^r \ket{\psi_0^{(e)}},
\end{equation}
by using Eqs.~(\ref{eq:III-UpUUp}) and (\ref{eq:III-CpPe}). The
advantage of this formulation is that the oracle $\mathcal{O}$ is used
on the subspace $\mathcal{H}^{V_n}$ unchanged, as it can be seen in
Eq.~(\ref{eq:III-cpp}).  As a consequence, the coin operator
$C''^{(+)}$ acts on the total Hilbert space $\mathcal{H}^{C_{n'}}
\otimes \mathcal{H}^{V_{n'}}$ as if two nodes were marked which differ
only in their last bits. Fig.~\ref{fig:double-hypercube} illustrates
the pair of marked vertices.  Intuitively, this is compensated in
Eq.~(\ref{eq:III-uupp}) by alternating $C''^{(+)}$ with a coin that
marks no vertices at all.

Next, we show that we can use the uniform superposition initial state
$\ket{\psi_0}$ as an initial state to the quantum walk if the
iterations are carried out according to the right hand side of
Eq.~(\ref{eq:III-uupp}). Let $X$ denote the quantum \textsc{not} gate,
$\sigma_X$, acting on the last qubit.  Clearly, we have $X
\ket{\psi_0^{(e)}} = \ket{\psi_0^{(o)}}$, and $[X, U^{(+)}] = [X,
U''^{(+)}] = 0$. Thus we can rewrite the desired initial state
(\ref{eq:psi0}) as $\ket{\psi_0} = (X+\Id)/\sqrt2 \ket{\psi_0^{(e)}} $
and see that
\begin{equation}
  (U^{(+)}U''^{(+)})^r \ket{\psi_0} = \frac1{\sqrt2} (X+\Id) 
  (U^{(+)}U''^{(+)})^r \ket{\psi_0^{(e)}}
\end{equation}
holds. In the right hand side we can discover Eq.~(\ref{eq:III-uupp})
which yields the state $\ket{\xtg'}$ with $1 - \Or(1/n')$ probability,
where $\xtg'$ is the image of $\xtg$ by the map (\ref{eq:map}). This
probability is distributed uniformly between the image $\xtg'$ and the
anti-image $\xtg' \oplus 1$ due to the multiplication by
$(X+\Id)/\sqrt{2}$.

\bibliographystyle{apsrev}

\end{document}